# High-Performance $MoS_2$ Field-Effect Transistors Enabled by Chloride Doping: Record Low Contact Resistance (0.5 kΩ·µm) and Record High Drain Current (460 µA/µm)

Lingming Yang[1], Kausik Majumdar[2*], Yuchen Du[1], Han Liu[1], Heng Wu[1], Michael Hatzistergos[3], Py Hung[2], Robert Tieckelmann[2], Wilman Tsai[4], Chris Hobbs[2], and Peide D. Ye[1#]

[1]*School of Electrical and Computer Engineering, Purdue University, West Lafayette, IN 47906, U.S.A.*
[2]*SEMATECH, Albany, NY 12203, U.S.A.* [3]*SUNY CNSE, Albany NY 12203, U.S.A.* [4]*Intel Corporation, Santa Clara, CA 95054, U.S.A.*
Tel: 1-765-494-7611, Fax: 1-765-496-7443, E-mail: [#]yep@purdue.edu, [*]kausik.majumdar@sematech.org

### *Abstract*

In this paper, we report a novel chemical doping technique to reduce the contact resistance ($R_c$) of transition metal dichalcogenides (TMDs) – eliminating two major roadblocks (namely, doping and high $R_c$) towards demonstration of high-performance TMDs field-effect transistors (FETs). By using 1,2 dichloroethane (DCE) as the doping reagent, we demonstrate an active n-type doping density > $2\times10^{19}$ $cm^{-3}$ in a few-layer $MoS_2$ film. This enabled us to reduce the $R_c$ value to a record low number of 0.5 kΩ·µm, which is ~10× lower than the control sample without doping. The corresponding specific contact resistivity ($\rho_c$) is found to decrease by two orders of magnitude. With such low $R_c$, we demonstrate 100 nm channel length ($L_{ch}$) $MoS_2$ FET with a drain current ($I_{ds}$) of 460 µA/µm at $V_{ds}$ = 1.6 V, which is twice the best value reported so far on $MoS_2$ FETs.

### *Introduction*

Semiconducting TMDs possess unique electrical and optical properties due to their d-electron orbitals and 2D nature [1,2]. Among TMDs, $MoS_2$ has attracted the most attention for its potential applications in low-power electronics [3,4]. However, high $R_c$ value limits the device performance of $MoS_2$ FETs significantly and the realization of ohmic contacts for $MoS_2$ remains a challenge so far [5]. There were several attempts to reduce $R_c$ including use of low workfunction metal [6] and employing edge contact concept [7]. One of the keys to resolve this issue is to dope the $MoS_2$ film, however doping the atomically thin film is nontrivial and requires a simple and reliable process technique [8-10]. In this work, we demonstrate such a doping technique enabling high-performance $MoS_2$ FET.

### *Fabrication and Physical Characterization*

Fig. 1(a) schematically shows the $MoS_2$ back-gate FET fabricated in this work. Few-layer $MoS_2$ flakes were mechanically exfoliated from bulk $MoS_2$ on a 90 nm $SiO_2/p^{++}$ Si substrate and then soaked in DCE. Acetone and isopropanol rinses were used to remove the residue of the chemical. After e-beam lithography, Ni (30 nm)/Au (60 nm) were deposited to form S/D contacts. The thickness of the $MoS_2$ flake was identified by the optical image (Fig. 2(a)) and measured by the AFM (Fig. 2(b)). The flake thickness was ~4 nm, corresponding to about 6 monolayers. Fig. 2(c) shows an SEM image of a fabricated TLM structure. The presence of Cl in DCE treated $MoS_2$ film was confirmed by XPS and SIMS, as shown in Fig. 3 (a) and (b). In Fig. 4, we observe a relative blue shift in the binding energies of the core level peaks of the $MoS_2$ sample that was treated with DCE, which results from an upward shift in the Fermi level, and hence can be attributed to an n-type doping of the sample. However, we note that Cl, when acts as an adatom dopant, results in p-type doping in $MoS_2$ film [11]. Thus, such n-type doping can be attributed to the donation of extra electron when substitution of $S^{2-}$ by $Cl^-$ takes place, particularly at the sites of sulfur vacancies in the $MoS_2$ film.

### *Contact Resistance Reduction*

The TLM resistances of $MoS_2$ FETs at 50 V back-gate-bias ($V_{bg}$) with and without the Cl doping are plotted as a function of contact separations in Fig. 5(a). The extracted $R_c$ is significantly reduced from 5.4 kΩ·µm to 0.5 kΩ·µm after the Cl doping. Such an improvement in $R_c$ is attributed to the doping induced thinning of tunneling barrier width. In Fig. 6, we observe that the extracted $R_c$ is a weak function of temperature (although the sheet resistance changes by a factor of 2), indicating the dominance of tunneling component of the current over thermionic component at the contact interface. In order to determine the $\rho_c$, the transfer lengths ($L_T$) of Ni-$MoS_2$ junctions are extracted by the TLM and are determined to be 60 nm and 590 nm for the contacts with and without the Cl doping, respectively. Compared with the control sample without the Cl doping, the $\rho_c$ is reduced from $3\times10^{-5}$ $\Omega\cdot cm^2$ to $3\times10^{-7}$ $\Omega\cdot cm^2$ when the DCE treatment time is 36 hours, as shown in Fig. 7. The n-type doping concentration ($N_d$) by chloride is ~$2.3\times10^{19}$ $cm^{-3}$ extracted from the slope of the TLM fitting when $V_{bg}$ is 0 V. Fig. 8 shows the channel resistance and the $R_c$ as a function of $V_{bg}$ for a 1 µm device. Usually in back-gated $MoS_2$ FETs, $R_c$ strongly depends on $V_{bg}$ because $V_{bg}$ would electrostatically dope the semiconductor underneath the contact, thus reducing the $R_c$. In this work, the $R_c$ shows very weak dependence on $V_{bg}$ when $V_{bg}$ is larger than -30 V, indicating heavily doped S/D regions are realized. Since back gate is not necessary for achieving the low $R_c$ any more, it paves the way to realize three-terminal top-gate low-$R_c$ $MoS_2$ FETs. The present Cl doping technique with DCE treatment is also valid for the other TMD materials such as $WS_2$, whose $E_F$ is pinned near the middle of the band bandgap.

### *Electrical Performance of $MoS_2$ FET*

Fig. 9 shows the output characteristics of 100 nm $L_{ch}$ $MoS_2$ FETs with and without the Cl doping. The reduced $R_c$ helps to boost the $I_{ds}$ from ~ 110 µA/µm to 460 µA/µm at $V_{ds}$ = 1.6 V, which is twice of the best reported value so far on $MoS_2$ FETs at the same $L_{ch}$ [6]. Fig. 10 shows the components of total resistance ($R_{total}$) indicating mitigation of the adverse dominance of high Schottky S/D contact resistance ($R_{sd}$) at 100 nm $L_{ch}$. Such reduction in $R_{sd}$ also results in excellent current saturation, as observed in Fig. 9. The transfer characteristics of the two devices are shown in Fig. 11. Due to its relatively large bandgap and ultra-thin channel, we achieved an excellent $I_{on}/I_{off}$ of ~$6.3\times10^5$. Considering the thick gate oxide (90 nm) used in this work, the $I_{on}/I_{off}$ ratio can be further improved by EOT scaling down. As shown in Fig. 12, the intrinsic long channel field-effect motility ($\mu_{FE}$) as a function of gate electric field is calculated for different $L_{ch}$ by appropriately eliminating the $R_{sd}$ effect with a peak $\mu_{FE}$ of 50-60 $cm^2$/Vs. Fig. 13 benchmarks the $I_{ds}$ ($V_{ds}$ = 1.6 V) and the $R_c$ for $MoS_2$ FETs in literature [5-7, 12-13]. Due to the significant reduction of $R_c$, the present work shows superior performance at various $L_{ch}$ compared with existing literature. These results indicate that the Cl doping by the DCE treatment is an effective way to realize low contact resistance $MoS_2$ FETs. Table 1 summarizes the electrical performance of the presented devices.

### *Conclusion*

For the first time, a record low $R_c$ of 0.5 kΩ·µm is achieved on the $MoS_2$ FET with Cl doping technique. As a result, the $\rho_c$ significantly decreases from $3\times10^{-5}$ $\Omega\cdot cm^2$ to $3\times10^{-7}$ $\Omega\cdot cm^2$. The 100 nm $L_{ch}$ $MoS_2$ FETs show a record high $I_{ds}$ of 460 µA/µm at $V_{ds}$ = 1.6 V, which is twice of the best reported $I_{ds}$ on any TMD FETs. As a result, this technique is promising for realizing high-performance top-gate low-$R_c$ $MoS_2$ FETs as well as other TMD based electronic devices.

### *References*

[1] B. Radisavljevic *et al.*, *Nature Nanotechnology*, **6**, p. 147, 2011 [2] H. Wang, *et al.*, *IEDM*, P. 88, 2012 [3] Y. Yoon *et al.*, *Nano Letters*, **11**, p. 3768, 2011 [4] K. Majumdar *et al.*, *EDL*, **35**, p. 402, 2014 [5] H. Liu *et al.*, *ACS Nano*, **8**, p. 1031, 2012 [6] S. Das *et al.*, *Nano Letters*, **13**, p. 3396, 2013 [7] W. Liu, *et al.*, *IEDM*, p. 400, 2013 [8] Y.C. Du *et al.*, *EDL*, **34**, p. 1328, 2013 [9] H. Fang *et al.*, *Nano Letters*, **12**, p. 3788, 2012 [10] H. Fang *et al.*, *Nano Letters*, **13**, p. 1991, 2013 [11] J. Chang, *et al.*, *arXiv: 1305.7162*, 2013 [12] Y.C. Du *et al.*, *EDL* (in press), 2014 [13] J. Lee, *et al.*, *IEDM*, p. 491, 2013

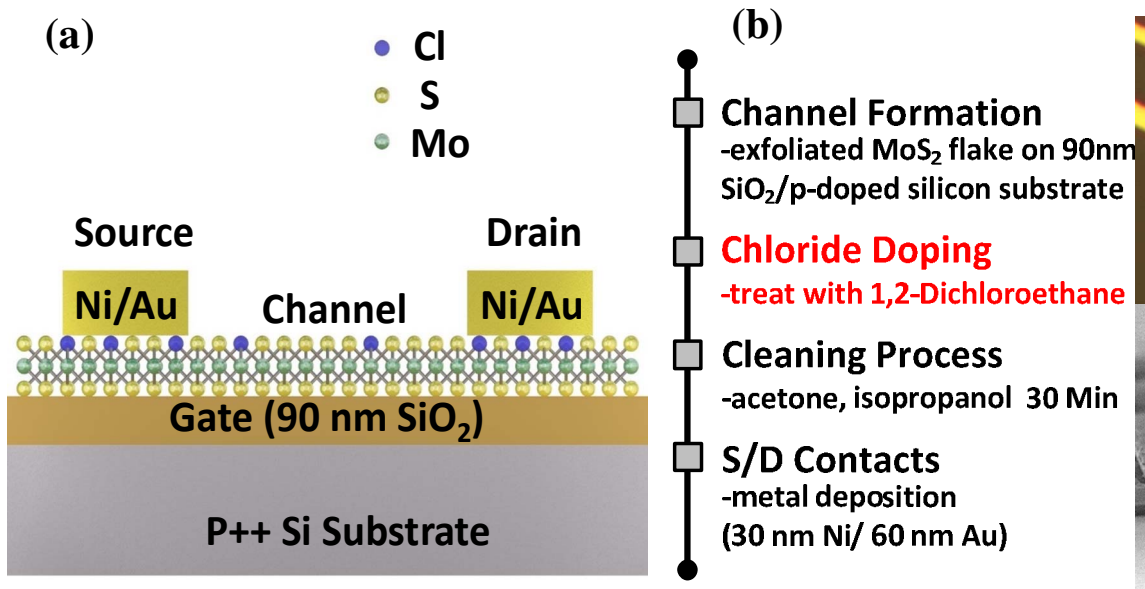

**Fig. 1** (a) Schematic of the $MoS_2$ back-gate FET fabricated in this work. The gate dielectric is 90 nm $SiO_2$. The S/D contact metal is Ni (30 nm)/Au (60 nm). (b) Process flow for the $MoS_2$ back-gate FETs with the exfoliated $MoS_2$ flakes.

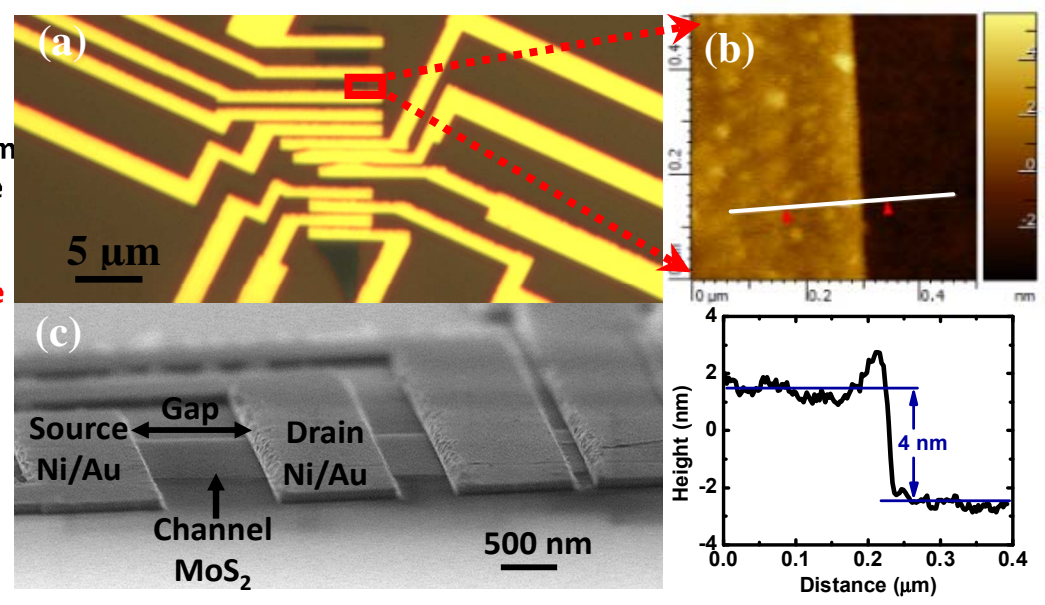

**Fig. 2** (a) Optical image of the few-layer $MoS_2$ FETs with the Ni/Au contacts on 90 nm $SiO_2/p^{++}$ Si substrate. (b) AFM image of a ~4 nm thick $MoS_2$ flake and the measured height at the flake edge. (c) SEM image of a $MoS_2$ TLM structure. The scale bar is 500 nm.

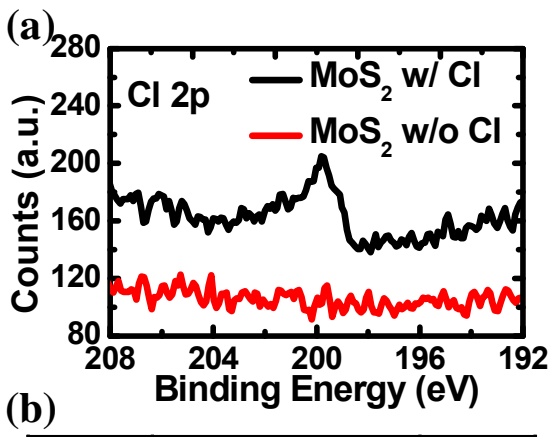

(b)

| Cl isotope | SIMS count (a. u.) | | Ratio |
|---|---|---|---|
| | w/ Cl | w/o Cl | |
| $^{35}Cl$ | 2451330 | 146257 | 16.76 |
| $^{37}Cl$ | 750188 | 33215 | 22.59 |

**Fig. 3** Cl signal from $MoS_2$ after the doping confirmed by (a) XPS and (b) SIMS.

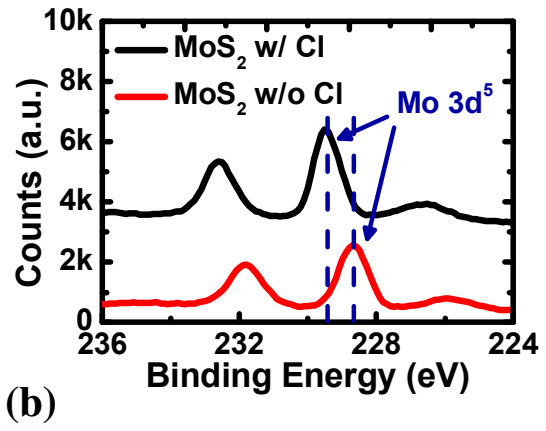

(b)

| Peak | Binding energy (eV) | | Shift (eV) |
|---|---|---|---|
| | w/o Cl | w/ Cl | |
| Mo $3d^5$ | 228.66 | 229.42 | + 0.76 |
| S $2p^3$ | 161.39 | 162.22 | + 0.83 |

**Fig. 4** (a) XPS spectra of Mo $3d^5$ w/ and w/o the Cl doping. A blue shift of 0.76 eV is observed. (b) Binding energy of the core levels w/ and w/o the Cl doping.

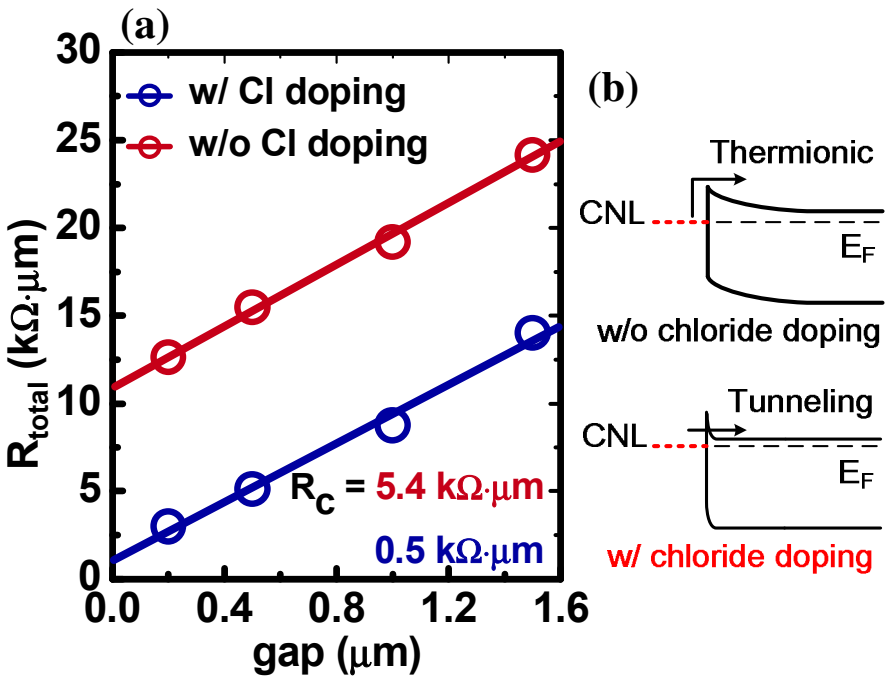

**Fig. 5** (a) TLM resistances of $MoS_2$ FETs w/ and w/o the Cl doping at $V_{bg}$ = 50 V. The $R_c$ is reduced from 5.4 kΩ·µm to 0.5 kΩ·µm (b) Band diagram of the metal-$MoS_2$ contacts w/ and w/o the Cl doping. $R_c$ is reduced due to the doping induced thinning of tunneling barrier width.

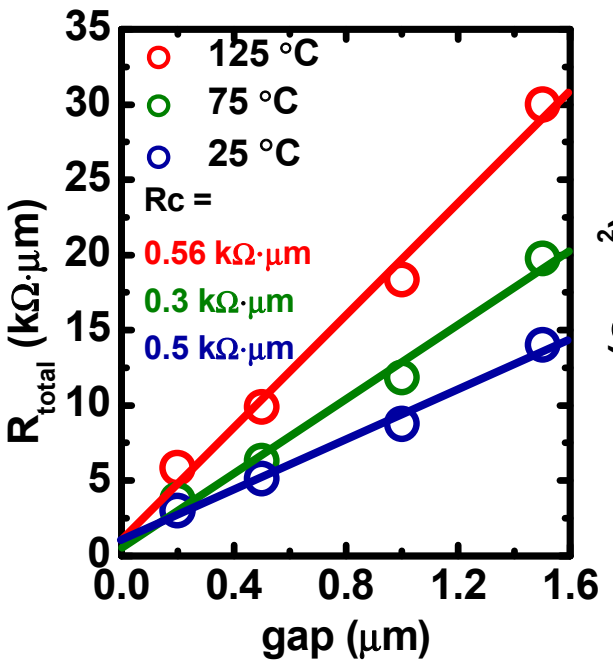

**Fig. 6** TLM resistances of $MoS_2$ FETs at 25, 75, and 125 ºC. The extracted $R_c$ remains similar from 25 ºC to 125 ºC, indicating the current is dominated by tunneling.

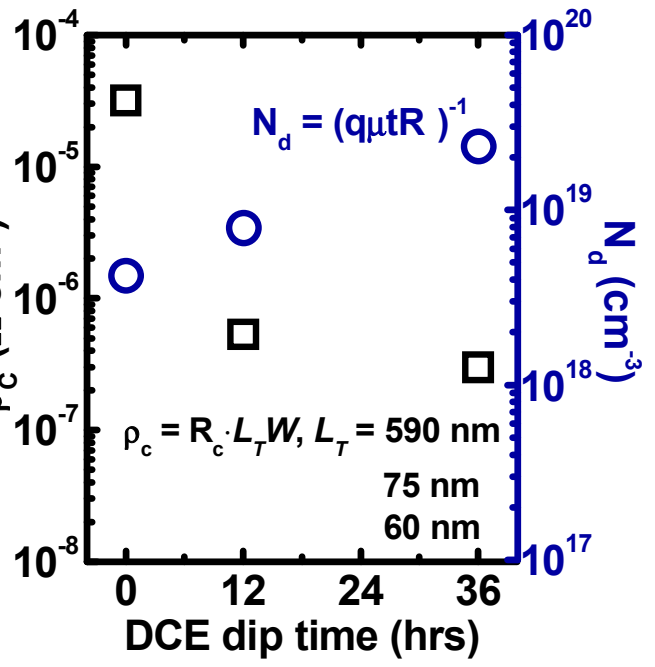

**Fig. 7** DCE dip time dependence of $\rho_c$ and $N_d$ with the DCE treatment. The $\rho_c$ is reduced by 100 after a 36 hours DCE treatment. N-type doping density of $2\times10^{19}$ $cm^{-3}$ is

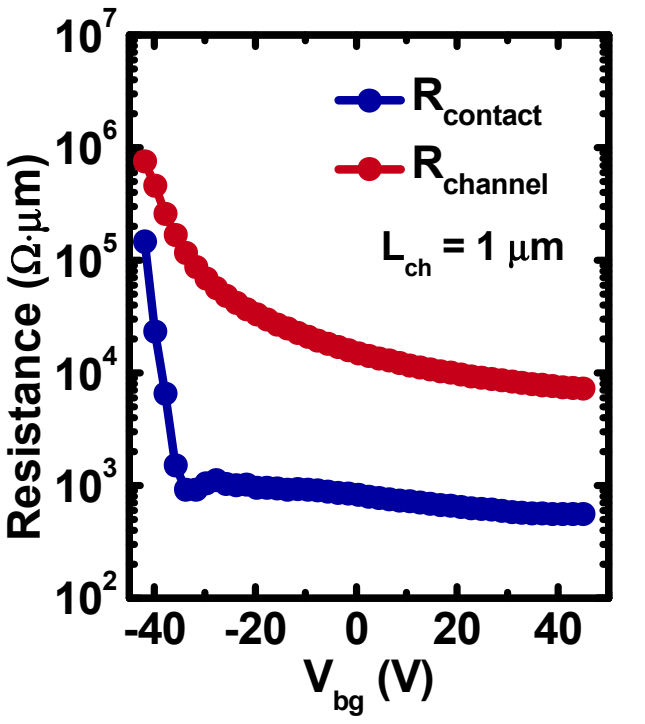

**Fig. 8** $R_{channel}$ and $R_{contact}$ vs. $V_{bg}$ for the 1 µm device. The $R_{contact}$ shows very weak dependence on the $V_{bg}$ when $V_{bg}$ > -30 V indicating heavily doped contact regions.

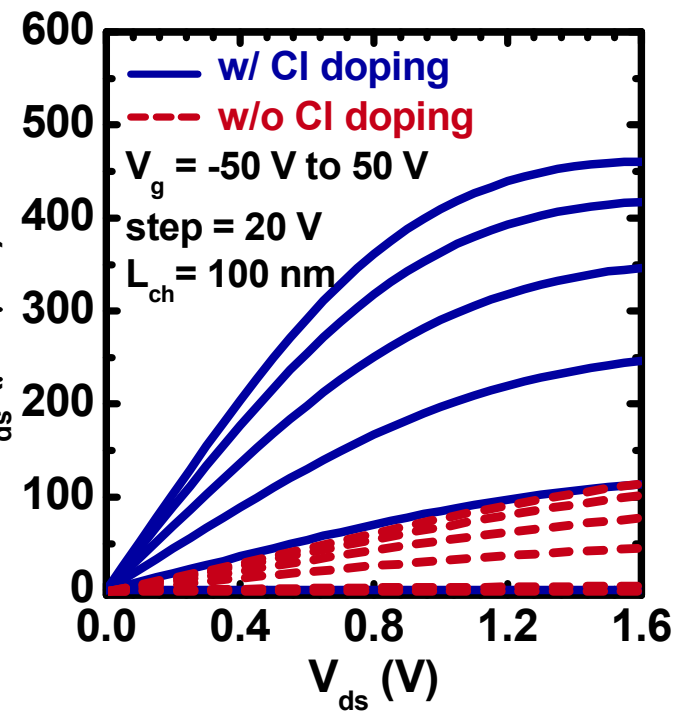

**Fig. 9** Output characteristics of the 100 nm $L_{ch}$ $MoS_2$ FETs w/ and w/o the Cl doping. A record high $I_{ds}$ of 460 µA/µm is obtained.

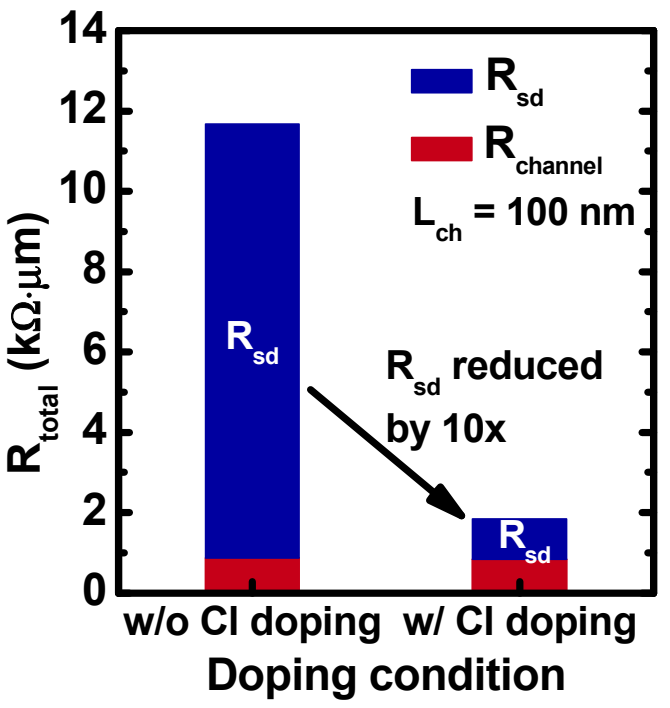

**Fig. 10** Component of $R_{total}$ for the two devices. The $R_{total}$ of 100 nm $L_{ch}$ $MoS_2$ FET is reduced from 11.7 kΩ·µm to 1.85 kΩ·µm due to the Cl doping.

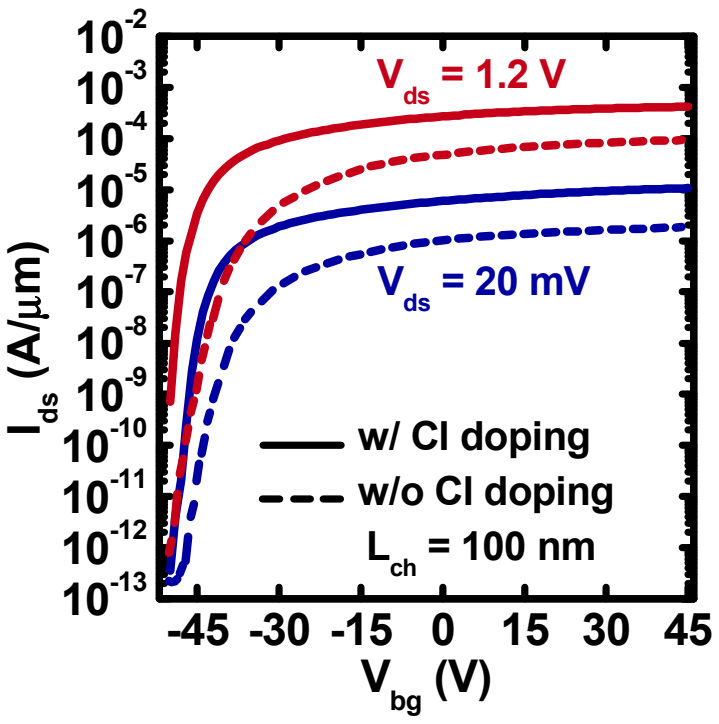

**Fig. 11** Transfer characteristic curves of the 100 nm $L_{ch}$ $MoS_2$ FETs w/ and w/o the Cl doping. The $I_{on}/I_{off}$ ratio is $6.3\times10^5$ at $V_{ds}$ = 1.2 V.

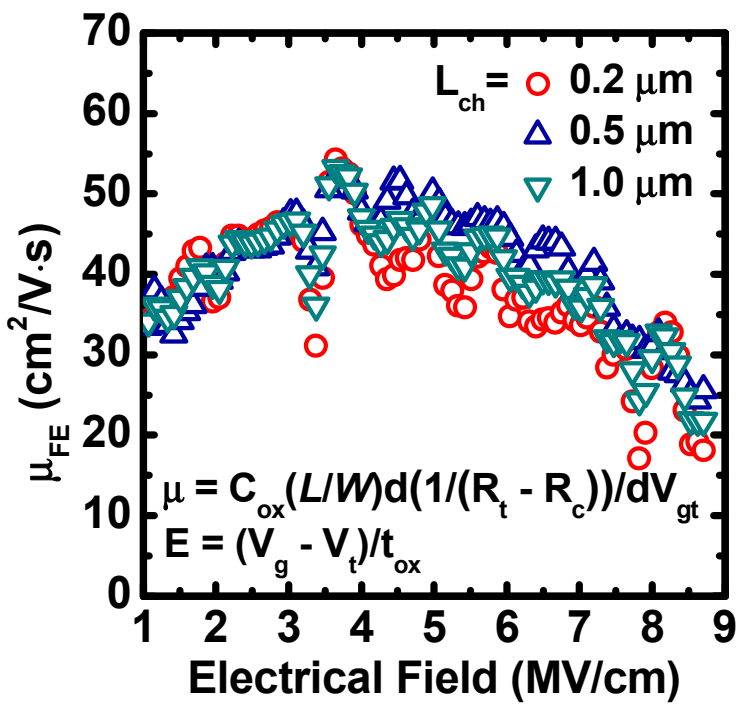

**Fig. 12** Calculated intrinsic field-effect motilities as a function of the *gate field* for the $MoS_2$ FET with various $L_{ch}$.

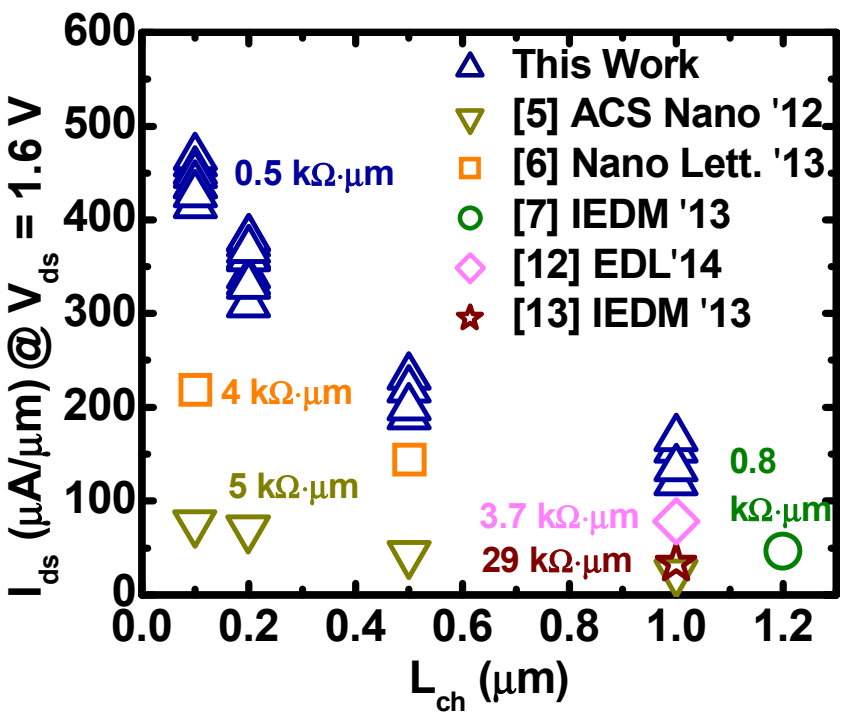

**Fig. 13** Benchmarking of the $I_{ds}$ @ $V_{ds}$ = 1.6 V and the $R_c$ in the reported $MoS_2$ back-gate FETs.

| Summary of present work | |
|---|---|
| EOT | 90 nm |
| $L_{ch}$ | 100 nm |
| $R_c$ | 0.5 kΩ·µm |
| $\rho_c$ | $3\times10^{-7}$ Ω·cm² |
| $I_{ds}$ | 460 µA/µm @ 1.6 V |
| $I_{on}/I_{off}$ | > $6.3\times10^5$ @ 1.2V |
| max. $\mu_{FE}$ | 50-60 cm²/V·S |

**Table. 1** Summary of the electrical performance of the $MoS_2$ FETs in this work.

***Acknowledgement:***

The work at Purdue University is supported by SEMATECH and SRC. The authors would like to thank Hong Zhou, Yexin Deng and Zhe Luo for the valuable discussions and technical assistance.